\documentclass[a4paper,10pt,onecolumn]{article}

\usepackage{graphicx}     
\usepackage{subfigure}    
\usepackage{indentfirst}  %
\usepackage{amsmath}      
\usepackage{natbib}       
\newcommand{\D}{\displaystyle}

\title{Physical Foundation of the Mathematical Concepts
       in the Nonstandard Analysis Theory of Turbulence }
\author{Feng Wu\\Department of Mechanics and Mechanical
Engineering,\\ University of Science and Technology of China, Hefei
230026, China}

\date{}  

\begin{document}

\newcommand{\supercite}[1]{\textsuperscript{\cite{#1}}}

\maketitle

\begin{abstract}
\noindent The physical foundation of the main mathematical concepts
in the nonstandard analysis theory of turbulence are presented and
discussed. The basic fact is that there does not exist the absolute
zero fluid-volume. Therefore, the corresponding physical object to
the absolute point is just the uniform fluid-particle. The
fluid-particle, in general, corresponds to the monad. The uniform
fluid-particle corresponds to the uniform monad, the nonuniform
fluid-particle to the nonuniform monad. There are two kinds of the
differentiations, one based on the absolute point, another based on
the monad. The former is adopted in the Navier-Stokes equations, the
latter in the fundamental equations, the closed forms are the
equations (7)-(11) in this article, in the nonstandard analysis
theory of turbulence. The continuity of fluid is shown by virtue of
the concepts of the fluid-particle and fluid-particle in lower
level. The character of the continuity in two cases, the standard
and nonstandard analysis, is presented in the paper. And the
difference in the discretization between the Navier-Stokes equations
and the equations (7)-(11) is pointed out too.


\end{abstract}

In papers \cite{fir} \cite{sec}   \cite{thr}, one new description of
turbulence is presented and the preliminary discussion on some
important and key concepts about the description is given too. The
new description of turbulence is called as Nonstandard Analysis
Theory of Turbulence(NATT). In this article, we will have a
discussion on the reasonability of the basic mathematical concepts
in the NATT. Some of these concepts are in being in the standard
analysis, but they have new meanings in the NATT. And the others are
not in being in the standard analysis at all.

Obviously, the reasonability of these mathematical concepts
originates in the physical foundation of them. The mathematical
logic is important, but should not be first.In the mathematical
logic,the introduction of these concepts should only meet a demand
for being not contradictory.

These mathematical concepts are: the point, the structure of point,
the point-average, the infinitesimal, the monad, the new
derivative,the continuity of fluid, etc.. We will discuss their
physical meanings in order to show the reasonability of these
mathematical concepts as follows.

\section{Point and continuity of fluid }
A fluid is composed of numerous moleculae, but could be thought of
as continuous provided that the scales of any physical process in
the fluid are much greater than the free path of molecule. Then the
flow of the fluid usually is called as a flow field, i.e., the fluid
in motion is modeled as a flow field. The flow field is continuously
composed of the real number points, and is called as real number
space mathematically. Yet after contemplation it is obvious that
there does not exist the ``point" of real number space in a real
fluid. Therefore, the physical meaning of the point of the real
number space( i.e., flow field), or in other words, the physical
reality from which  the concept of point is drawn, should be
indicated. In fact, the logic of the process,that the real number
space( i.e., flow field) is drawn from the flow fluid composed of
moleculae, should be as follows.

It is well known that the motion of the moleculae contained in a
very small volume will reach, by collision with each other, the
equilibrium in a very small time interval. By virtue of this fact,
the concept of the fluid-particle has been presented. The fluid, in
fact, should be thought to be physically composed of the
fluid-particles. The fluid-particle is a very small volume of fluid
and is real physical object. Hence, the concept of the mathematical
point is drawn from the fluid-particle, i.e., the abstraction of the
fluid-particle is the point. After the abstraction from the
fluid-particles, the flow field composed of the points is obtained.
So the point forming flow field is an abstracted concept and is
drawn from the fluid-particle, that is to say, the concept of point
is based on the fluid-particle.

Obviously, it is necessary to make this abstraction for knowing the
properties of flow. The mathematical derivation could hardly be put
on the base of the fluid-particle. Otherwise, on the base of real
number space composed of the real number points the mathematical
derivation, by which the law and properties of fluid flow can be
obtained, is made reasonably.

The fluid-particle is a very small volume of fluid. However small
the scale of the fluid-particle is , it is still finite from the
angle of physical practice. We would like to point out some features
about the scale of the fluid-particle.

The physical world is possessed of the hierarchical structure. The
scales of the fluid-particles in different levels are different.
Surely, the scale of the fluid-particle in higher level should be
much greater than that in lower level. The scale of the
fluid-particle in the lowest level in a physical problem should not
be less than a certain scale $\xi$. The certain scale $\xi$ is the
lowest scale in the scales, any one of which can form the volume
which contains enough number of moleculae for stable statistical
average value of physical quantities. The fluid volume, the linear
scale of which is less than the scale $\xi$, does not have certain
average value of physical quantities. Therefore, that fluid volume
could not be called as a fluid-particle.

But the scale of a fluid-particle in any level could not be too
large to show the variation of physical quantities with time and
space.

It is important to indicate that the scales of the fluid-particles
in any level are all objective. The scale of the fluid-particle is
not determined by people at will, but by the nature of the specific
physical problem under study. In other words, the scale of the
fluid-particle should be put on the base of physics and be
corresponding to the nature of the specific physical problem under
discussion. As an example, the scale of the fluid-particle in
atmospheric flow is much greater than that in the flow around the
wing of an aircraft.

In practical activities, people usually can not give how large, in
the concrete, the fluid-particle is, yet we believe that the scale
of the fluid-particle is not set by people's will, but is objective.

By the concept of the fluid-particle, the clearer physical meaning
of the continuity of a fluid can be given, that is:  If there is a
space scale of $\xi$, enough number of moleculae are contained in
the fluid volume with linear scale $\xi$, so that the statistical
averaging over the motion of these moleculae can give the stable
value, i.e., the fluid-volume with linear scale $\xi$ just is a
fluid-particle in certain level. And, the space scales of any
physical process in the fluid are all greater than the scale $\xi$.
Further, the motion of the moleculae contained in the volume with
linear scale $\xi$, by collision with each other, has already
reached thermodynamic equilibriums in the time interval less than or
equal to the least time scale in the various time scales of the
various physical processes in the fluid. Then the fluid can be
thought of as continuous. In this case, the effect of fluid composed
of moleculae can be ignored, and the real flow of fluid will be
modeled after the flow field.

\section{Uniform and non-uniform points }
The fundamental physical idea in the NATT is that the fluid-particle
is wholly uniform in a laminar flow, but not uniform, i.e.,there is
internal structure in the fluid-particle in the case of turbulence.
By that mentioned above, though the fluid-particle in the turbulence
can be divided at will into smaller and uniform fluid-particles in
mathematical abstract, we could not reject this idea, because the
fluid-particle can not be divided at will in physical practice. The
scale of the fluid-particle has the objective physical base in
practice.

What is the mathematical meaning of this fundamental physical idea
in the NATT?

As is stated above, a point(the absolute geometric point) should be
mathematically drawn from only the uniform fluid-particle. If the
non-uniform fluid-particle was, in mathematical abstraction, thought
of as an absolute geometric point too, the internal structure of the
non-uniform fluid-particle can not be described. So this will go
against the physical reality. Thus, the non-uniform fluid-particle
should be, in mathematical abstraction, taken as the monad. The
linear scale of the monad is infinitesimal
$\varepsilon$,$\varepsilon\neq 0$,but $\varepsilon>0$. The internal
structure can be permitted in the monad. From the nonstandard
analysis, the infinitesimal $\varepsilon$ is also number, i.e., the
nonstandard number. The real number is called as the standard
number.

Even in the case of a laminar flow, in fact, the monad is also the
mathematical abstraction of the uniform fluid-particle, but is
uniform, no internal structure in the monad. The monad no having
internal structure essentially is identified with the absolute
point. This concept can be shown clearly by the figure 1-2.

\begin{figure}[ht]
 \centering
  \includegraphics[width=1.0\textwidth]{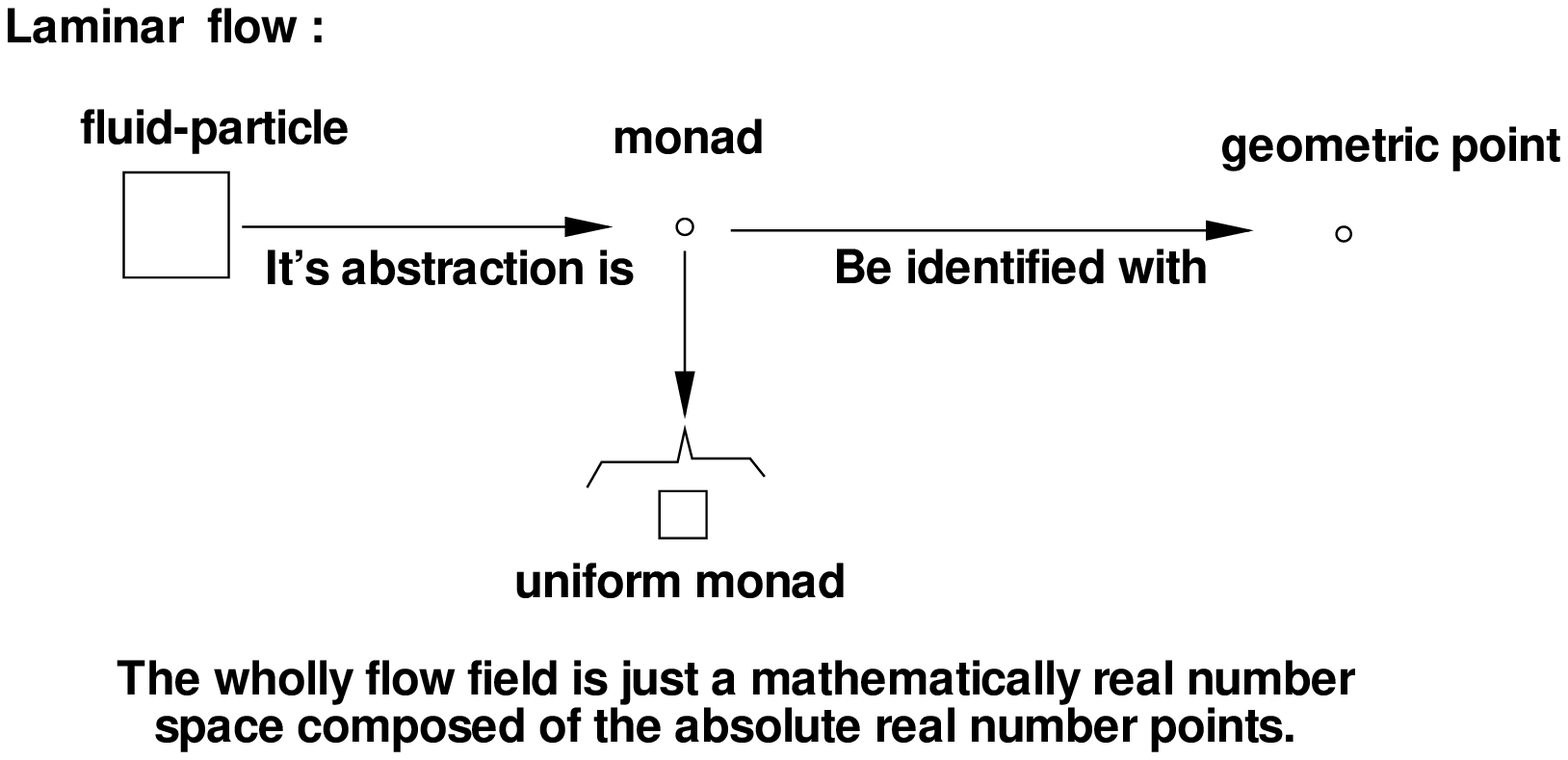}
  \caption{Mathematical-abstraction of the fluid-particle in laminar flow }
  \label{fig1fluid}
\end{figure}

\begin{figure}[ht]
  \centering
  \includegraphics[width=1.0\textwidth]{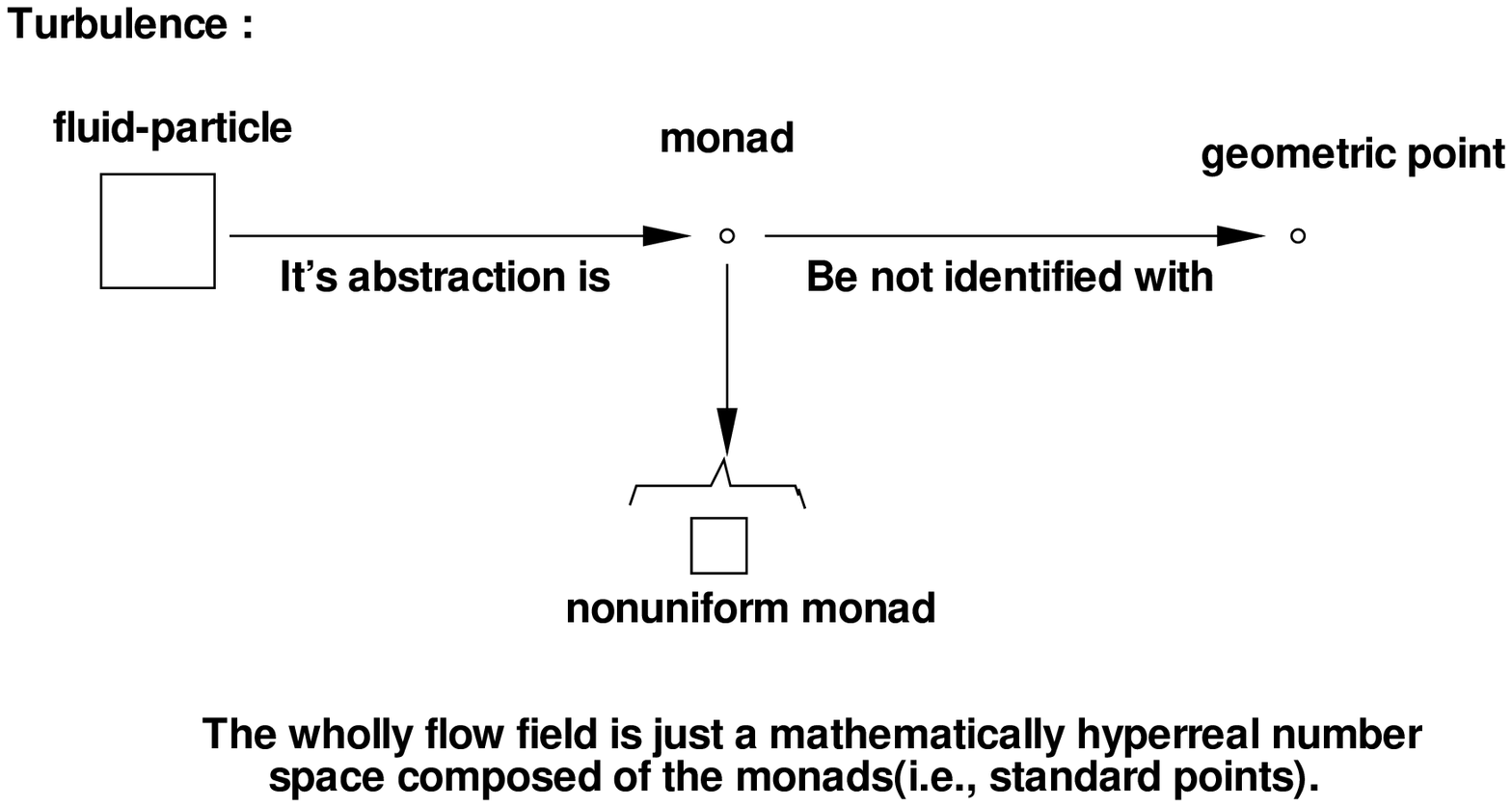}
  \caption{Mathematical-abstraction of the fluid-particle in turbulence }
  \label{fig2fluid}
\end{figure}

Therefore, the physical meaning of mathematical monad is the
fluid-particle, i.e., the monad is the mathematical abstraction of
the fluid-particle. In the case of a laminar flow the fluid-particle
is wholly uniform, so the monad(mathematical abstraction of the
fluid-particle) is also uniform. And yet, in turbulence, the
fluid-particle is not wholly uniform, so the abstract monad is
non-uniform. In paper \cite{fir}, the monad is also called as the
standard point being different from the absolute point. Thus,
correspondingly there are also uniform and non-uniform standard
points(usually called as uniform and non-uniform points for short).
In a laminar flow, the uniform standard point drawn from the uniform
fluid-particle is identified essentially with the absolute point. So
the physical meaning of the absolute point is also the uniform
fluid-particle.

The structure of the point in NATT represents just the internal
structure of the standard point, rather than the absolute point,
which has no the internal structure. Obviously, the internal
structure of the standard point is the mathematical abstraction
just from the internal structure of the fluid-particle. So the
physical meaning of the point-structure is just the internal
structure of the fluid-particle.

Finally, it should be pointed out that the fluid-particle is the
physical reality and has finite scale. Because the scale of the
fluid-particle is very small, it is,in the mathematical
abstraction, thought of as monad, the scale of which is
infinitesimal, even thought of as the absolute point, the scale of
which is zero.

In paper \cite{fir}, it was shown that the fluid-particle, whether
or not is uniform, is composed of numerous fluid-particles in lower
level(Fig.3), and a fluid-particle in lower level is uniform and
contains still enough number of fluid moleculae for the stable
statistic-average.

\begin{figure}[ht]
\centering
  \includegraphics[width=1.0\textwidth]{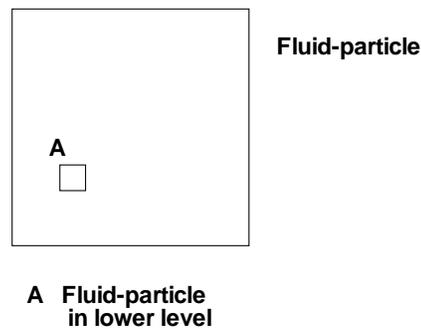}
  \caption{Fluid-particle composed of fluid-particles in lower level }
  \label{fig3fluid}
\end{figure}

The mathematical abstract of the fluid-particle in lower level is
the nonstandard point(i.e., the internal point of a monad).
Therefore, the nonstandard point(i.e., the internal point of a
monad) is uniform point.

So a turbulent field is possessed of hierarchical structure: the
global turbulent field, composed of the monads, i.e., the
abstraction from the fluid-particles, is in a level. And the monad
field(i.e., the abstraction from the flow in the fluid-particle),
composed of the nonstandard points(the mathematical abstraction from
the fluid-particles in lower level), is in another lower level.

\section{Two kinds of differentiations}

Having a function $f(x)$, the derivative of which is

\begin{equation}
\frac{\partial f}{\partial x}=\lim_{\triangle x\rightarrow
0}\frac{f(x+\triangle x)-f(x)}{\triangle x}
\end{equation}

Here the mathematical meaning of $\triangle x\rightarrow 0$ is that
$\triangle x$ tends to the absolute zero. But, in many cases, people
can not give the clearly physical meaning of the absolute zero, for
example, the physical meaning of ``a fluid volume tends to the
absolute zero". In many cases, in fact, there is not ``absolute
zero" in physics. The physical meaning of the statement of ``a fluid
volume tends to the absolute zero" is ambiguous. Therefore, the
statement of that $\triangle x$ tends to absolute zero(i.e.,
$\triangle x\rightarrow 0$) in mathematics does not correspond to
the fact that any physical quantity tends to absolute zero, but
corresponds to, for example in the case of fluid flow, the fact that
physical quantity $\triangle x$ tends to the scale of the uniform
fluid-particle. In other words, the mathematical abstraction of the
fact that $\triangle x$ tends to the scale of the uniform
fluid-particle in physics is that $\triangle x\rightarrow 0$. The
physical meaning of $\triangle x\rightarrow 0$ is that $\triangle x$
tends to the scale of the uniform fluid-particle.

In NATT, there is other derivative:
\begin{equation}
\frac{\partial f}{\partial
x}=\frac{f(x+\varepsilon)-f(x)}{\varepsilon}
\end{equation}

Here $\varepsilon$ is nonstandard number infinitesimal. The
$\varepsilon$ is not arbitrary infinitesimal, but the linear scale
of a monad. The monad is the abstraction from the fluid-particle. So
the infinitesimal $\varepsilon$, the linear scale of the monad, is
logically the abstraction from the linear scale of the
fluid-particle, namely, the physical meaning of the infinitesimal
$\varepsilon$ is the linear scale of the fluid-particle.

The definition (2) has already been presented in the nonstandard
analysis. But there the infinitesimal is arbitrary, here in NATT the
$\varepsilon$ is certain infinitesimal, the linear scale of a monad.

The definition (1), in general, denotes mathematically that the
$\frac{\partial f}{\partial x}$ is the rate of increment of function
$f$ at some point(absolute geometric point), i.e., the slope of
tangent to the curve of function $f$. But, in many cases, the
absolute point does not exist in physics, let alone the rate of
increment of function at this point. In the case of fluid, it can be
understood that the physical meaning, which the definition (1)
corresponds to, is the increment(i.e., the increment of the physical
quantity $f$ between two neighbour uniform fluid-particles in some
level) divided by the scale of the fluid-particle in the same level.
Let A,B be two neighbour uniform fluid-particles, the scale of the
fluid-particle be $L$, then the physical meaning of $\frac{\partial
f}{\partial x}$ in definition (1) is that

 $$\frac{f(B)-f(A)}{L}$$

Therefore, from the angle of physics, the definition (1) is
applicable to only the uniform fluid-particle in some level.
Correspondingly, in mathematics, the definition (1) is applicable to
only the uniform point in the same level.

The right side of the definition (2), from the angle of mathematics,
denotes the increment(i.e., the increment of function $f$ between
two corresponding internal points of two infinite close monads)
divided by the linear scale of the monad. And the physical meaning
of the right side of definition (2) is the increment(i.e., the
increment of physical quantity function $f$ between two
corresponding fluid-particles in lower level, which are uniform in
NATT, of two neighbour fluid-particles) divided by the scale of the
fluid-particle. Let $L$ be the linear scale of the fluid-particle,
$b$,$a$ be the two corresponding fluid-particles in lower level of
two neighbour fluid-particles $B$,$A$, i.e.,$b$,$a$ are located at
the same place in respective fluid-particles(Fig.4).

\begin{figure}[ht]
  \centering
  \includegraphics[width=1.0\textwidth]{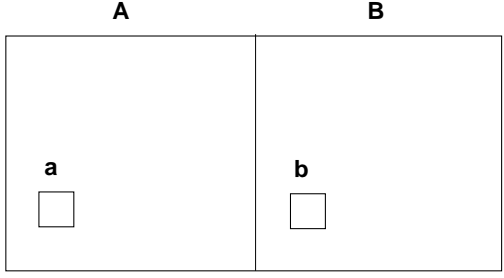}
  \caption{Positions of a  and b  in fluid-particles $A$ and $B$ }
  \label{fig4fluid}
\end{figure}

So the meaning of the definition (2) is that $\frac{\partial
f}{\partial x}$ in (2) is the abstraction from the physical quantity
$\frac{f(b)-f(a)}{L}$. The reasonability of the mathematical
abstract $\frac{\partial f}{\partial x}$ in (2) is based on the real
physical quantity $\frac{f(b)-f(a)}{L}$. The definition (2) can be
applicable to both the uniform and nonuniform fluid-particle in
physics, or to both the uniform and nonuniform point in mathematics.

If, in fluid mechanics, the conservation laws of physics(the
conservation of mass, momentum and energy) are applied to uniform
and nonuniform fluid-particle in some level respectively, the
different results will be obtained. In the case of uniform
fluid-particle, the mathematical abstraction of the results obtained
is the Navier-Stokes equations(for the incompressible fluid)

\begin{equation}
\frac{\partial U_{i}}{\partial x_{i}}=0
\end{equation}

\begin{equation}
\frac{\partial U_{i}}{\partial
t}+\frac{\partial(U_{i}U_{j})}{\partial
x_{j}}=-\frac{1}{\rho}\frac{\partial P}{\partial
x_{i}}+\nu\nabla^{2}U_{i}
\end{equation}

The equations (3)(4) hold for only the uniform point. In the case of
the nonuniform fluid-particle, the mathematical abstraction of the
results obtained is the equations(for the incompressible fluid)
\begin{equation}
\frac{\partial\widetilde{U_{i}}}{\partial x_{i}}=0
\end{equation}

\begin{equation}
\frac{\partial\widetilde{U_{i}}}{\partial
t}+\frac{\partial\widetilde{U_{i}U_{j}}}{\partial
x_{j}}=-\frac{1}{\rho}\frac{\partial\widetilde{P}}{\partial
x_{i}}+\nu\nabla^{2}\widetilde{U_{i}}
\end{equation}

Here the sign ``$\sim$" denotes the average over all the nonstandard
points contained in a monad.

The equations (5)(6) hold for both the uniform and nonuniform point.
The equations (5)(6) will degenerate into the Navier-Stokes
equations (3)(4) in the case of the uniform point.

The differentiation (1) is based on the limit, expressed as the
statement of $\delta-\varepsilon$, and in the frame of
$\delta-\varepsilon$. The differentiation (2) is not based on the
limit, and out of the frame of $\delta-\varepsilon$.

The statement of $\delta-\varepsilon$ shows mathematically the
endless process of infinite tending, and makes the expression of the
process stricter mathematically. But it, in the stating the process
of infinite tending, for example infinite tending to zero, does not
give any room to physics, i.e., it shows the process of tending to
the absolute zero or to the absolute nil. However, the real physics
should be: the process of infinite tending to zero is just a
mathematical abstraction from the real process of gradually tending
to the scale of the uniform fluid-particle in some level.

The statement of $\delta-\varepsilon$ obliterates this real uniform
fluid-particle. In many cases, this obliteration does not have an
effect on the results. Yet the obliteration will make some problems
sometimes. How if the fluid-particle is not uniform?  The structure
in the fluid-particle can not be shown by the absolute point.
Turbulence just is the case. Here the statement of
$\delta-\varepsilon$ should be abandoned and the differentiation (2)
is adopted. The differentiation (2) is mathematically based on the
infinitesimal. In the sixties of the last century, the nonstandard
analysis was presented. It is proved strictly in the nonstandard
analysis that there exists the infinitesimal. This gives the
mathematical foundation to the new description of turbulence.

\section{continuity}

In section 1, we showed the physical meaning of the fluid
continuity. From the physical reality the mathematical model of
fluid, i.e., continuous flow field, is obtained. In the continuous
flow field, the physical properties(i.e., the various physical
quantities on all points of the field) vary continuously with time
and space. Yet the mathematical description of the continuously
varying of the physical quantities is different in two cases, the
standard and nonstandard analysis.

In the case of standard analysis, when the function representing the
physical quantity is continuous mathematically, the physical
quantity is thought of as continuous. In other words, the physical
quantity varies continuously with time and space, the function $f$
representing the physical quantity is, in mathematics, continuous
function. The definition of continuous function is that(Fig.5)
\begin{figure}[ht]
  \centering
  \includegraphics[width=1.0\textwidth]{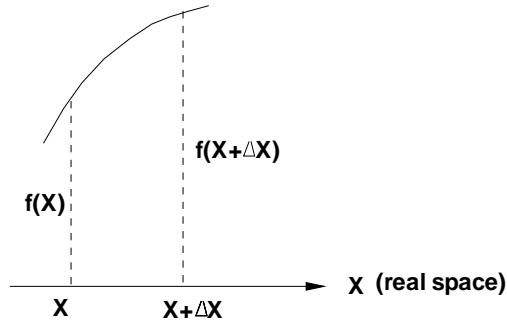}
  \caption{Continuous function $f(x)$ }
  \label{fig5fluid}
\end{figure}

$$\lim_{\triangle x\rightarrow 0}[f(x+\triangle x)-f(x)]=0$$

Note that the function $f(x)$ is characterized by only the function
value.

But, in nonstandard analysis, this is not the case. Here the
function representing the physical quantity should be written as
$f(x,x')$,  the sign $x$ denotes the $x$-monad in some hyper real
number space, $x'$ is the internal coordinates in the monad, i.e.,
$x'$ is the coordinates of the internal point of the
$x$-monad(Fig.6). The $f(x,x')$ expresses the value of the physical
quantity at the internal point $x'$ of the $x$-monad. Then the
continuously varying, with time and space, of the physical quantity
expressed by the function $f(x,x')$ means that not only the value
but also the structure of the function $f(x,x')$ vary continuously.
In other words, what is continuous function of $f(x,x')$ means that
not only the increment of $f(x+\triangle x,x')-f(x,x')$ tends to
infinitesimal, but also the shape of the function $f(x+\triangle
x,x')$ tends infinitely to the shape of the function $f(x,x')$, when
$\triangle x$ tends to infinitesimal $\varepsilon$(the linear scale
of a monad), i.e., the two monads(the monad $(x+\triangle x)$ and
the monad $(x)$ ) are infinitely close to each other(Fig.7).
\begin{figure}[ht]
  \centering
  \includegraphics[width=1.0\textwidth]{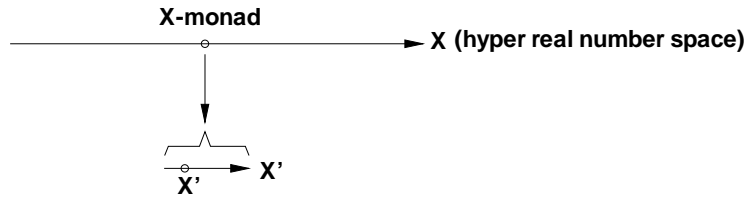}
  \caption{The internal point $x'$ of $x$-monad }
  \label{fig6fluid}
\end{figure}

\begin{figure}[ht]
  \centering
  \includegraphics[width=1.0\textwidth]{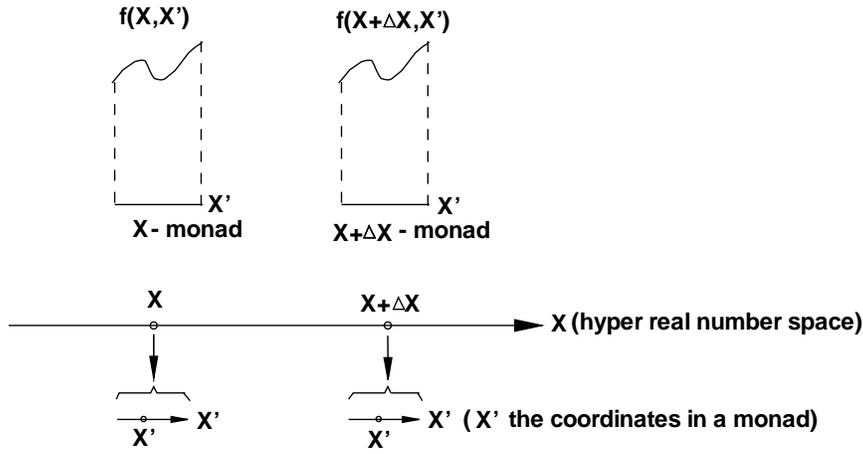}
  \caption{Continuous function $f(x,x')$ }
  \label{fig7fluid}
\end{figure}

Here the function $f(x,x')$ is possessed of both the value and the
shape(i.e., the variation of the function $f(x,x')$ with $x'$ in
$x$-monad). So the variation of not only the function value but also
the function shape should be continuous, if the physical quantity
function is continuous. This means that the continuity of the fluid
should be the continuity of the function shape as well as the
function value. There are two sides in the continuity of fluid in
the case of the nonstandard analysis.

In the standard analysis, the physical meaning of the mathematically
continuity of the function expresses the relationship between only
two neighbour uniform fluid-particles. When people say that the
function representing the physical quantity is continuous, the
physical meaning of the continuity is that the difference of the
value of the physical quantity between two neighbour uniform
fluid-particles is very small.

Similarly, in the nonstandard analysis, the mathematical concept of
continuity depicts, in physics, just the relationship between two
neighbour fluid-particles(both uniform and nonuniform). The
difference of the physical quantity, its value and structure(i.e.,
the structure of the physical quantity function), between the two
neighbour fluid-particles is very small. In the NATT, this is called
as assumption of close property between two neighbour
fluid-particles, in abstraction, between two infinitely close
monads. The assumption is just the continuity(physically
corresponding to the case of both uniform and nonuniform
fluid-particle) in the nonstandard analysis, the expansion of the
continuity(physically corresponding to the case of only uniform
fluid-particle) in normally the standard analysis. The reasonability
of this assumption should be checked up by the reasonability of the
results from the assumption.

There is discontinuity in opposition to the continuity. In order to
show these concepts clearly, the concept of the fluctuation in a
fluid will be discussed on further as follows.

If the precision of the measurement in turbulence is very high, are
the data obtained from the measurement  continuous(no fluctuation)?

As is stated above, the concept of the monad is drawn from the
fluid-particle. Here, the monad and fluid-particle is in the space
of four dimensions, time and space. What is the fluid-particle in
space(three dimensions) is clear. Yet the meaning of the
fluid-particle in time-direction needs to be indicated. Let the time
scale of the fluid-particle be $\delta_{t}$, the time scale of the
fluid-particle in lower level be $\tau$. Surely, the scale of
$\delta_{t}$ and $\tau$ are all objective, and determined by the
physical nature of the problem under discussion. The scale $\tau$
needs to meet the requirements of the conditions as follows: The
motion of the fluid moleculae contained in the space fluid-particle
in lower level, by collision with each other, has already reached
thermodynamic equilibrium in the time period of $\tau$. The
time-direction scale, which is infinitesimal, of the monad in four
dimensions and the time-direction scale, which is high order
infinitesimal, of the monad-interior point(nonstandard point) in
four dimensions are the mathematical abstraction of $\delta_{t}$ and
$\tau$ respectively.

It is known that if the measurement in a fluid is carried out in the
level of atom and molecule, i.e., the measurement of the motion of
atom and molecule, the results of the measurement will show the
random motion, controlled by the quantum law, of the atom and
molecule. This measurement is out of the studying at present. Now
the objective of the measurement is the mean quantity of the motion
of numerous moleculae.

Assume that the motion of numerous moleculae contained in the space
fluid-particle in lower level, by collision with each other, has
already reached the thermodynamic equilibrium in only the time
interval greater than or equal to $\tau$. The sampling time in the
measurement is $t_{0}$. If $t_{0}<\tau$, the data of the measurement
can not indicate the physical properties of this fluid particle in
lower level, but the average value, which is instable, of the motion
properties of the moleculae contained in the fluid volume being less
than the fluid-particle in lower level. So the data are fluctuant.
If $t_{0}>\tau$, but $\ll\delta_{t}$(the time scale of the
fluid-particle), the data of measurement should be continuous(not
fluctuation) and obey the Navier-Stokes equations. The measurement,
in fact, is carried out in the range of time scale of a
fluid-particle. But the usual measurement in turbulence are not the
cases.

In the turbulence measurement at present, the time-samples are taken
from the numerous time scales of the numerous fluid-particles in
lower level, which are contained in the various fluid-particles of
four dimensions. The results for one sample are just the statistic
average values of the motion properties, which are reached
equilibrium by molecule-collision with each other in the interval
$\tau$, of the numerous moleculae contained in the space
fluid-particle in lower level. In other words, in the mathematical
abstract, the samples are taken from the numerous time-scales of the
numerous internal points of various monads(Fig.8). The essential
character of this measurement is that the measurer can not know
which internal point of a monad is measured, but know only which
monad contains the measured internal point. So the measurer know
only the range over a monad, but can not know the strict position of
the measured objective. This is the uncertainty of turbulence
measurement. Therefore, obtained data will show the fluctuation
provided that the motion properties of different internal points of
a monad are obviously different.
\begin{figure}[ht]
  \centering
  \includegraphics[width=1.0\textwidth]{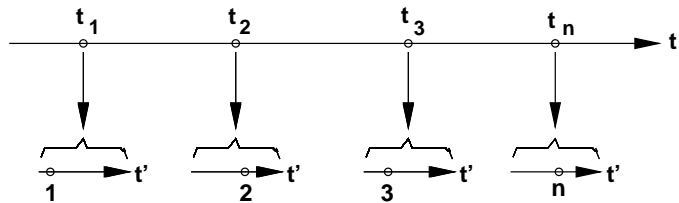}
  \caption{The scheme of the sampling in turbulence measurement }
  \label{fig8fluid}
\end{figure}

Though it is possible in theory that the measurements are carried
out continuously on the different internal points of one monad, this
precision can be reached by modern technology, the continuous data
can be obtained in the monad. And the consecutive measurement is in
the next monad. And so on and so forth, there are continuous
measurements in various monads one by one. However, the concrete
operation of the measurement is borne hardly at all, because it is
nearly the process in which infinite samples are taken. This process
of the measurement as well as the process of solving infinite
equations is not possible. Therefore, this measurement exists only
in theory and imagining, but not in reality.

Thus the practical measurement of turbulence will give, as mentioned
above, the fluctuant results. This fluctuation results from the
measurement of the regular flow(the turbulence is also regular).
Note that the fluid flow itself is one thing, the result from the
measurement of the fluid flow is another. They have relation closely
with each other, but are not the same thing. The irregular
measurement results does not denote the irregularity and
un-continuity of the fluid flow itself, conversely, the continuity
and regularity of the fluid flow does not exclude the possibility of
the irregular results(or called as the fluctuation) from the
measurement of the fluid flow. When the turbulence measurement is
carried out, what is met with is the fact that the irregular and
fluctuant data are obtained from the measurement of regular and
continuous fluid flow. It is maybe said that the
turbulence-fluctuation, in fact, just is the fluctuation of the data
from the turbulence measurement.

About the fluctuation, in a word, except the random results of
measurement in molecule level, there are the fluctuation results
from not reaching the stable state of statistic average, i.e., from
the measurement in the range of the scale less than that of the
fluid-particle in the least level, and the fluctuation results
especially from the uncertainty of turbulence measurement(in the
level of global field). Only the turbulence-measurement in a monad,
the mathematical concept drawn from the fluid-particle, possibly
gives the continuous results.

In NATT, there is a mathematical concept of the
monad-average(point-average). The physical meaning of the
point-average is the average over the motion of numerous lower level
fluid-particles in a fluid-particle. The point-average is drawn from
the average over the fluid-particle. Here the point-average is
different from the normal average in turbulence researching, and it
is not over measurement data. Surely, we easy connect the
point-average in NATT with the normal average in turbulence
researching. The value of the point-average can be obtained by the
theoretical computation(after discretization of the controlling
equations the point-average is, in fact, the grid-average), then the
average value, being comparable with the average-value from the
measurement, is easy obtained from the grid-average.

The standard and nonstandard analysis are mentioned in NATT. In a
laminar flow, the standard analysis is employed; in turbulence, the
nonstandard analysis is employed. Either laminar or turbulent flow,
in fact, either the standard or nonstandard analysis all can be
employed. The standard analysis means that there is description of
one level, the nonstandard analysis means that there is description
of multi-levels. In the standard analysis, the description of a
fluid flow is based on the uniform fluid-particle(i.e., the
fluid-particle in a laminar flow, the fluid-particle in lower level
in a turbulent flow). The fluid-particle as well as the
fluid-particle in lower level is wholly uniform in the case of a
laminar flow, so the fluid-particle can be thought of as basic
element in a laminar flow. And just the fluid-particle corresponds
to the physical practice. Therefore, the standard analysis, the
description of one level, is employed in a laminar flow. However, in
turbulence, the fluid-particle is not wholly uniform and the
fluid-particles in lower level contained in the fluid-particle are
possessed of respective different motion properties. Obviously,
there are multi-levels in turbulence. This is why the nonstandard
analysis is employed in turbulence.  The standard analysis, in
theory, can also be employed, i.e., the description of one level and
still the Navier-Stokes equations controlling the flow, in
turbulence, but the fluid-particle in lower level must be the basic
element. Yet the fluid-particle in lower level does not correspond
to the physical practice. There does exist some unsuitability in
practice. The evident example of this unsuitability is that the
difficulty of the very enormous amount of calculation work will be
met, as the Navier-Stokes equations are solved by numerical
calculation in turbulence.

\section{Discretization of the equations}

There are, for the incompressible fluid, three sets of the closed
equations for a turbulent flow in the NATT.

A.Choice one:
\begin{equation}
\frac{\partial\widetilde{U_{i}}}{\partial
x_{i}}=0,\quad\frac{\partial\widetilde{U_{i}}}{\partial
t}+\frac{\partial(\widetilde{U_{i}}\widetilde{U_{j}})}{\partial
x_{j}}=-\frac{1}{\rho}\frac{\partial\widetilde{P}}{\partial
x_{i}}+\nu\nabla^{2}\widetilde{U_{i}}+o(\varepsilon^{2})
\end{equation}

B.Choice two:
\begin{equation}
\frac{\partial U_{i}}{\partial x_{i}}=0,\quad \frac{\partial
U_{i}}{\partial t}+\frac{\partial(U_{i}U_{j})}{\partial
x_{j}}=-\frac{1}{\rho}\frac{\partial P}{\partial
x_{i}}+\nu\nabla^{2}U_{i}
\end{equation}
\begin{equation}
\frac{\partial u_{i}}{\partial x_{i}}=0,\quad\frac{\partial
u_{i}}{\partial t}+U_{j}\frac{\partial u_{i}}{\partial
x_{j}}+u_{j}\frac{\partial U_{i}}{\partial
x_{j}}-2\frac{\partial(u_{i}u_{j})}{\partial
x_{j}}=-\frac{1}{\rho}\frac{\partial p}{\partial
x_{i}}+\nu\nabla^{2}u_{i}+o(\varepsilon^{3})
\end{equation}

C.Choice three:
\begin{equation}
\frac{\partial\widetilde{U_{i}}}{\partial
x_{i}}=0,\quad\frac{\partial\widetilde{U_{i}}}{\partial
t}+\frac{\partial(\widetilde{U_{i}}\widetilde{U_{j}})}{\partial
x_{j}}+\frac{\partial(u_{i}u_{j})}{\partial
x_{j}}=-\frac{1}{\rho}\frac{\partial\widetilde{P}}{\partial
x_{i}}+\nu\nabla^{2}\widetilde{U_{i}}+o(\varepsilon^{3})
\end{equation}
\begin{equation}
\frac{\partial u_{i}}{\partial x_{i}}=0,\quad\frac{\partial
u_{i}}{\partial t}+\widetilde{U_{j}}\frac{\partial u_{i}}{\partial
x_{j}}+u_{j}\frac{\partial \widetilde{U_{i}}}{\partial
x_{j}}=-\frac{1}{\rho}\frac{\partial p}{\partial
x_{i}}+\nu\nabla^{2}u_{i}+o(\varepsilon^{3})
\end{equation}

Here the $\varepsilon$ is the linear scale of a monad and is
infinitesimal(nonstandard number). The $U_{i}$,$P$ are the
instantaneous velocity and pressure. The $u_{i}$,$p$ are the
fluctuant velocity and pressure. They are the functions of
$(x_{1},x_{2},x_{3},t,x^{\prime}_{1},x^{\prime}_{2},x^{\prime}_{3},t^{\prime})$.
The $\widetilde{U_{i}}$,$\widetilde{P}$ are the
point-averages(monad-averages) of the instantaneous velocity and
pressure. The independent variables of the
$\widetilde{U_{i}}$,$\widetilde{P}$ are only
$(x_{1},x_{2},x_{3},t)$. Note that here the fluctuant quantities are
different from the normal fluctuations in a turbulent measurement.
The definition of the fluctuations $u_{i}$,$p$ is that
$u_{i}=U_{i}-\widetilde{U_{i}}$ and $p=P-\widetilde{P}$. Obviously,
the $u_{i}$,$p$ are defined in a monad. Therefore, the $u_{i}$,$p$
would be called as the fluctuations in a monad, or the internal
fluctuations.

The differentiation (2) has been applied to
$\frac{\partial}{\partial t}$,$\frac{\partial}{\partial x}$ in these
equations. So the equations can be applied to both the uniform and
nonuniform monads. What is the physical meaning of the
discretization of the equations (7)-(11)? By virtue of the fact that
the infinitesimal $\varepsilon$ in the definition (2) is the linear
scale of a monad, the discretization of the equations (7)-(11) means
the coarsening of the monad. But the discretization of the
Navier-Stokes equations means, in the case of turbulence, the
coarsening of the uniform internal point, essentially identified
with the absolute point, of a monad. There are different meanings
between two. The monad is abstract from the fluid-particle, but the
scale of grid of discretization of (7)-(11) is, in general, not
equal to the scale of fluid-particle in practice. If the physical
relation between the neighbour grids is very close to the real
physical relation between the neighbour fluid-particles, the results
of computation will be approximately correct, and the discretization
is reasonable. Conversely, if the physical relation between the
neighbour grids is not close to that between the neighbour
fluid-particles, the correct results can not be obtained and the
discretization is not reasonable.

We know by that mentioned above that the Navier-Stokes equations are
based on the limit($\triangle x\rightarrow 0$,$\triangle
t\rightarrow 0$), and the Navier-Stokes equations can be applied to
only the uniform point(i.e., the abstract from the uniform
fluid-particle). Yet the equations of (7)-(11) can be permitted to
apply to the nonuniform point(i.e., the abstract from the nonuniform
fluid-particle). In turbulence, the fluid-particle, the abstraction
of which is the monad, is not uniform, but the fluid-particle in
lower level, the abstraction of which is the nonstandard point(i.e.,
the internal point of the monad), is uniform. Therefore, in
turbulence, the Navier-Stokes equations can be applied to only the
internal points of a monad rather than to the monad itself. But the
equations (7)-(11) can be applied to the monad itself. When the
computation of these equations is carried out, the discretization of
the Navier-Stokes equations means the coarsening of the internal
point of a monad, physically corresponding to the fluid-particle in
lower level, and the discretization of equations (7)-(11) means the
coarsening of the monad itself, physically corresponding to the
fluid-particle. Therefore, in the turbulent computation, the grid of
discretization corresponds to the fluid-particle in lower level in
the case of computation of the Navier-Stokes equations, but
corresponds to the fluid-particle itself in the case of computation
of the equations (7)-(11). In other words, only the very fine grids,
which introduces the very enormous amount of calculation work, can
meet the needs for obtaining the reasonable solution of the
Navier-Stokes equations, but the coarse grids can meet the needs for
the reasonable solution of the equations (7)-(11) in the case of the
turbulence computation. It does not need to be fine grids, so there
is not the difficulty of enormous amount of calculation work in the
turbulent computation of the equations (7)-(11).

First, the mathematical concepts are drawn from the physical
reality. And by virtue of the mathematical concepts a series of the
mathematical derivations and calculations are performed in order to
obtain the various results about the physical properties, for
example, the controlling equations. Then, after the discretization
of the controlling equations the computation of the
discretization-equations is carried out and the various physical
properties of the concrete problem are obtained. There are two
procedures, from the physical reality to the mathematical abstract
and from the mathematical abstract to the physical reality, the
latter is a converse procedure of the former.

\section{Conclusions}

In the sections above, the discussion on some important mathematical
concepts and the corresponding physical meanings to these concepts
is given. What are called as the corresponding physical meanings are
the physical objects from which the mathematical concepts are drawn.
The results are, in sum, as follows.


\hspace{0.0cm}MATHEMATICAL CONCEPTS\hspace{1.0cm}CORRESPONDING
PHYSICAL OBJECTS

$\ast$ \hspace{0.0cm}The absolute point\hspace{1.0cm}the uniform
fluid-particle

$\ast$ \hspace{0.0cm}The uniform point(uniform
monad)\hspace{1.0cm}the uniform fluid-particle

$\ast$ \hspace{0.0cm}The nonuniform point(nonuniform
monad)\hspace{1.0cm}the nonuniform fluid-particle

$\ast$ \hspace{0.0cm}The nonstandard point(internal point of a
monad)\hspace{1.0cm}the fluid-particle in lower level

$\ast$ \hspace{0.0cm}$\triangle x\rightarrow
0$\hspace{1.5cm}$\triangle x$ tends to the scale of uniform
fluid-particle

$\ast$ \hspace{0.0cm}$\D\frac{\partial f}{\partial
x}=\lim_{\triangle x\rightarrow 0}\frac{f(x+\triangle
x)-f(x)}{\triangle x}$\hspace{1.5cm}the difference of the $f$-values
between only two

\hspace{6.6cm}  neighbour uniform fluid-particles divided

\hspace{6.6cm} by the linear scale of the fluid-particle

$\ast$ \hspace{0.0cm}The infinitesimal
$\varepsilon$\hspace{1.0cm}the linear scale of the fluid-particle

$\ast$ \hspace{0.0cm}$\frac{\partial f}{\partial
x}=\frac{f(x+\varepsilon)-f(x)}{\varepsilon}$\hspace{1.5cm}the
difference of $f-$values between the corresponding two

\hspace{4.5cm} lower level fluid-particles of the neighbour
fluid-particles

\hspace{4.6cm}divided by the linear scale of the fluid-particle

$\ast$ \hspace{0.0cm}The N-S equations\hspace{1.0cm}the conservation
equations(i.e., the mass, momentum and energy

\hspace{4.5cm}equations) about only the uniform fluid-particle

$\ast$ \hspace{0.0cm}The fundamental equations in the
NATT\hspace{1.0cm}the conservation equations about both the

\hspace{0.7cm}(the closed forms are the equations\hspace{1.7cm}
uniform and nonuniform fluid-particle

\hspace{0.6cm}  (7)-(11) in this article)

$\ast$ \hspace{0.0cm}The standard analysis\hspace{1.0cm}the flow
fluid composed of only uniform fluid-particles,

\hspace{5.0cm} only one level

$\ast$ \hspace{0.0cm}The nonstandard analysis\hspace{1.0cm}the flow
fluid composed of the fluid-particles,

\hspace{5.6cm} multi-level analysis

$\ast$ \hspace{0.0cm}The discretization of the N-S
equations\hspace{1.0cm}the coarsening of the nonstandard point into

\hspace{2.0cm}in turbulence computation\hspace{1.5cm}the grid,
corresponding to the

\hspace{7.7cm}fluid-particle in lower level

$\ast$ \hspace{0.0cm}The discretization of the equations
(7)-(11)\hspace{0.80cm}the coarsening of the monad into the grid,

\hspace{2.2cm}in turbulence computation\hspace{1.7cm}corresponding
to the fluid-particle itself

$\ast$ \hspace{0.0cm}The continuity of fluid\hspace{1.0cm}the
physically description of fluid-continuity is in

\hspace{5.0cm}the last paragraph of the section $1$ in this article
$$$$

Finally, what is stated in this paper is that the reasonability of
the mathematical concepts, especially in physics, is based just on
their corresponding physical reality. Every reasonable mathematical
concept, even mathematical derivation method, must be possessed of
its physical foundation, i.e., is the abstract from the physical
object(physical reality). In physics, there does not exist the
mathematical concept broken away from the physical reality. People,
sometimes, ignore carelessly this fact, and are limited to and only
stay on the stage of the pure mathematics. Because getting
accustomed, for example, to the absolute point in mathematical
abstraction, some researchers usually do not pay attention to the
effect of the fact that there does not exist the absolute point in
physical reality on the description of physical problem such as the
turbulence. And the physical problem becomes very difficult to
comprehend due to this carelessness. But after the contemplation
over the physical meanings of these mathematical concepts, in other
words, over what are the real physical objects from which these
mathematical concepts are drawn, the difficult problem may be
suddenly enlightened.

\bibliography{dbofref}

\begin{thebibliography}{9}
   \bibitem[1]{fir} F.Wu, Nonstandard analysis picture of turbulence, preprint
   physics/0308012\\
   (lanl.arXiv),2004.
   \bibitem[2]{sec} F.Wu, Some key concepts in nonstandard analysis
   theory of turbulence, Chin. phys. lett., Vol.22,(2005)2604.
   \bibitem[3]{thr} F.Wu, A new approach to the theory of turbulence,
   Journal of shanghai university(English edition), Vol.9,(2005)189.
\end{thebibliography}

\end{document}